\documentclass[12pt,american,english, journal,draftclsnofoot,onecolumn]{IEEEtran}
\usepackage[T1]{fontenc}
\usepackage[latin9]{inputenc}
\usepackage{color}
\usepackage{float}
\usepackage{textcomp}
\usepackage{amsmath}
\usepackage{amssymb}
\usepackage{stackrel}
\usepackage{graphicx}
\usepackage{setspace}
\usepackage{esint}
\makeatletter

\floatstyle{ruled}
\newfloat{algorithm}{tbp}{loa}
\providecommand{\algorithmname}{Algorithm}
\floatname{algorithm}{\protect\algorithmname}

\usepackage{colortbl}
\usepackage{cite}
\usepackage{nomencl}
\usepackage{flushend}
\usepackage{stfloats}
\usepackage[margin=8pt,font=footnotesize,labelfont=md,labelsep=colon]{caption}

\@ifundefined{showcaptionsetup}{}{%
 \PassOptionsToPackage{caption=false}{subfig}}
\usepackage{subfig}
\makeatother

\usepackage{babel}
\addto\captionsamerican{\renewcommand{\algorithmname}{Algorithm}}

\begin{document}

\title{NOMA Made Practical: Removing the Receive SIC Processing through
Interference Exploitation}

\author{Abdelhamid Salem,\textit{\normalsize{} Member, IEEE}{\normalsize{},
}Xiao Tong, Ang Li,\textit{\normalsize{} Senior Member, IEEE} and
Christos Masouros, \textit{\normalsize{}Senior Member, IEEE}.\\
\thanks{Abdelhamid Salem,\textit{ and} Christos Masouros, are with the department
of Electronic and Electrical Engineering, University College London,
London, UK, (emails: \{a.salem, c.masouros\}@ucl.ac.uk).

Xiao Tong and Ang Li are with the School of Information and Communications
Engineering, Faculty of Electronic and Information Engineering, Xi\textquoteright an
Jiaotong University, Xi\textquoteright an, Shaanxi 710049, China (emails:17745169490@163.com
and ang.li.2020@xjtu.edu.cn).%
} }
\maketitle
\begin{abstract}
Non-orthogonal multiple access (NOMA) is a powerful transmission technique
that enhances the spectral efficiency of communication links, and
is being investigated for 5G standards and beyond. A major drawback
of NOMA is the need to apply successive interference cancellation
(SIC) at the receiver on a symbol-by-symbol basis, which limits its
practicality. To circumvent this, in this paper a novel constructive
multiple access (CoMA) scheme is proposed and investigated. CoMA aligns
the superimposed signals to the different users constructively to
the signal of interest. Since the superimposed signal aligns with
the data signal, there is no need to remove it at the receiver using
SIC. Accordingly, SIC component can be removed at the receiver side.
In this regard and in order to provide a comprehensive investigation
and comparison, different optimization problems for user paring NOMA
multiple-input-single-output (MISO) systems are considered. Firstly,
an optimal precoder to minimize the total transmission power for CoMA
subject to a quality-of-service constraint is obtained, and compared
to conventional NOMA. Then, a precoder that minimizes the CoMA symbol
error rate (SER) subject to power constraint is investigated. Further,
the computational complexity of CoMA is considered and compared with
conventional NOMA scheme in terms of total number of complex operations.
The results in this paper prove the superiority of the proposed CoMA
scheme over the conventional NOMA technique, and demonstrate that
CoMA is an attractive solution for user paring NOMA MISO systems with
low number of BS antennas, while circumventing the receive SIC complexity.

$\,$\end{abstract}


\section{Introduction}

Non-orthognal multiple access (NOMA) technique has received significant
attention very recently as a viable multiple access technique for
communication networks \cite{magazinDing,NOMAbyme1,NOMAbyme2}. In
NOMA the transmitter superimposes the users signals in same frequency,
time, and code domains while being able to resolve the signals in
the power domain. The users with poor channel conditions (weak users)
are allocated with high transmission power levels, while the users
with strong channel conditions (strong users) are allocated with low
power levels. At reception, the weak users detect their signals by
treating the other users\textquoteright{} signals as noise. On the
contrary, the strong users first decode the signals of the weaker
users, then they detect their own signals by removing the weaker users\textquoteright{}
signals using successive interference cancellation (SIC) \cite{magazinDing}.
This is a significant known limitation of NOMA, which poses impractical
symbol-by-symbol complexity. 

The efficiency of NOMA technique has been extensively investigated
in the literature. For instance, the results in \cite{Ding2} showed
that NOMA can achieve superior performance comparing with orthogonal
multiple access (OMA) schemes. The performance of NOMA was analyzed
in \cite{Ding3} based on the availability of the channel state information
(CSI) at the transmitter. In \cite{newmain} a power minimization
problem for two-users multiple-input-single-output (MISO)-NOMA systems
was formulated and solved. The results in this work showed that the
proposed NOMA approach can enhance the performance of MISO systems.
To maximize the fairness among the users in NOMA systems, an optimal
power allocation scheme has been considered in \cite{nomafairness}. 

Furthermore, in NOMA systems when number of users is large, the interference
in the system might be strong. This interference will lead to increase
the complexity and the processing delay at the receivers. More relevant
to this work, in order to reduce the interference, complexity and
processing delay, user pairing scheme has been proposed and considered
in the literature\cite{twousers1,userpairs,pair1,pair2}. In this
scheme, each two users (pair) share a specific orthogonal resource
slot and NOMA technique is implemented among the users in each pair.
User pairing scheme has been widely investigated in the literature.
The authors in \cite{userpairs} proposed a user pairing scheme in
which the network area is divided into two regions, near and far regions,
and each far user is paired with a near user. The results in \cite{userpairs}
explained that, the performance gain performed by NOMA over OMA can
be further improved by paring the users whose channel conditions are
more distinctive. In \cite{pair1} MU-MIMO NOMA systems was considered,
in which the users are paired and share the same transmit beam-forming
vector. Under this scenario, the superiority of MIMO-NOMA over MIMO-OMA
has been proved for a two-user paring scenario. The authors in \cite{pair2}
proposed a greedy-search based user pairing scheme in order to maximize
the achievable sum rate of NOMA system.

In parallel, constructive interference (CI) precoding has received
research interest in the last few years \cite{CI1,A,CI2,Alodeh}.
CI precoding is also a non-orthogonal transmission approach which
exploits the known interference to improve the system performance.
Based on the knowledge of the CSI and the users messages, the BS can
classify the multi-user interference as constructive and destructive.
The constructive interference can be defined as the interference that
can move the received symbol deeper in the detection region of the
constellation point of interest. Accordingly, a constructive precoder
can be obtained to make the known interference in the system constructive
to the received symbols. The CI concept has been widely studied and
investigated in the literature. This line of research introduced in
\cite{CI1}, where the CI precoding technique has been proposed for
downlink MIMO systems, showing significant performance improvements
over conventional precoding. The first optimization based CI approach
was introduced in \cite{A} where a modified\emph{ }\textit{\emph{vector-perturbation}}
technique was proposed, in which the search of perturbing vectors
was limited to a specific area where the distances from the decision
thresholds are increased with respect to a distance threshold. In
\cite{CI2,Alodeh}, a symbol-level precoding scheme for downlink MU-MISO
system has been proposed. In these works the authors used the knowledge
of the CSI and data symbols to exploit the constructive interference
in the system.  Further work in \cite{Luxm1,Luxm2}, a general category
of CI regions has been considered, and the features of this region
have been studied. Different convex alphabet relaxation schemes for
vector precoding in MIMO broadcast channels have been proposed in
\cite{Rodrigo} to achieve interference-free communication over singular
channels. It has been shown in \cite{Muller} that vector precoding
can be implemented to reduce the transmission power of MIMO systems.
The authors in \cite{CI3,CI5} implemented CI precoding scheme in
wireless power transfer scenarios to minimize the total transmit power.
Recently in \cite{angLi,angliQAM} closed-form expressions of CI precoding
scheme for phase-shift keying (PSK) and quadrature amplitude modulation
(QAM) have been derived. These expressions have derived based on optimal
performance, thus its performance is equivalent to the optimization-based
CI schemes presented in the literature. Based on these closed form
expressions, in our previous works in \cite{Tcompaper,rspaper,errorratepaper,secrecypaper}
analytical expressions of the achievable sum-rate and error probability
of CI precoding have been derived for different scenarios. For more
details, the reader is referred to \cite{surveyAngLI} where the concept
of the CI precoding scheme and its practical implementation have been
presented and discussed in details.

In this work, we exploit the CI concept to address a major limitation
of NOMA systems. This is the need to apply SIC on a symbol-by-symbol
basis at the receiver, which introduces impractical complexity. Accordingly,
we propose an approach to entirely circumvent SIC, based on the concept
of CI. We introduce a new constructive multiple access NOMA (CoMA)
precoding technique that aligns the superimposed signals to the different
user equipments (UE) constructively to the signal of interest. The
key principle is shown in Fig. 1b, contrasting it with the classical
NOMA approach in Fig. 1a. Since for CoMA the superimposed signal aligns
with the data signal, the received symbol appears at the correct constellation
region, it does not require channel equalization and there is no need
to remove the interfering symbol at the receiver using SIC technique.
Critically, this new scheme allows the following key gains that make
NOMA practical:
\begin{enumerate}
\item A low complexity UE - by removing SIC from the receiver, CoMA allows
minimal receive signal processing as shown in Fig. 1b.\textcolor{blue}{{}
}
\begin{figure}
\noindent \begin{centering}
\textcolor{blue}{}\subfloat[\label{fig:1a-1}Conventional NOMA.]{\noindent \begin{centering}
\textcolor{blue}{\includegraphics[bb=50bp 150bp 880bp 450bp,clip,scale=0.55]{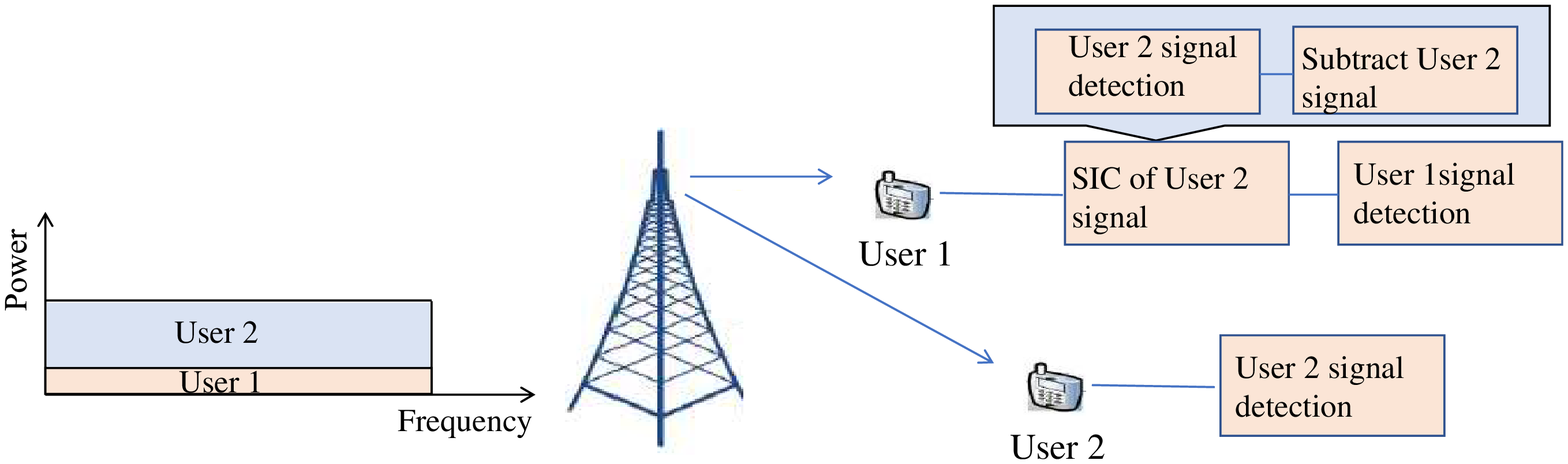}}
\par\end{centering}

\textcolor{blue}{}

}
\par\end{centering}

\noindent \begin{centering}
\textcolor{blue}{}\subfloat[\label{fig:1b-1}CoMA (Invention)]{\noindent \begin{centering}
\includegraphics[bb=0bp 200bp 900bp 520bp,clip,scale=0.55]{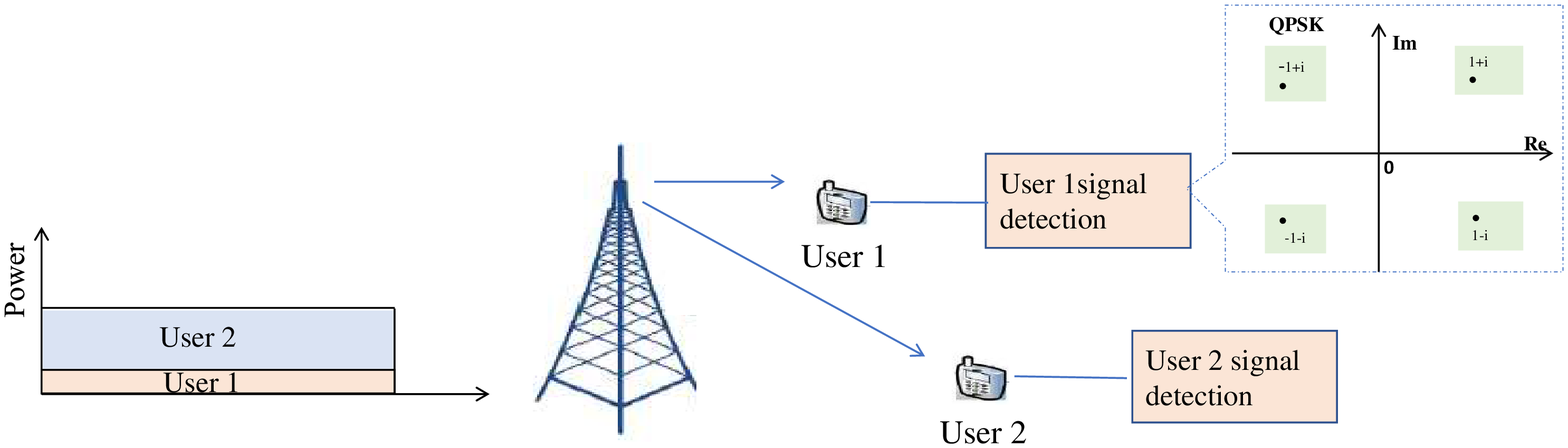}
\par\end{centering}

\textcolor{blue}{}}
\par\end{centering}

\textcolor{blue}{\protect\caption{\label{fig:1}Conventional NOMA and CoMA schemes.}
}
\end{figure}

\item Since channel equalization is not required, this removes the need
for channel state information (CSI) at the UE, which in turn\\
$\quad\quad\quad\quad\quad\text{\textbullet}$ Removes the overheads
associated with collecting and sharing CSI.\\
$\quad\quad\quad\quad\quad$\textbullet{} Removes the quantization
and noise-related errors in the CSI shared from the BS to each UE. 
\item Reduces the latency in processing the received signal on a symbol-by-symbols
basis at each UE.
\end{enumerate}
Due to the above key advantages, the proposed approach makes NOMA
more practical and fits different practical scenarios. In this regard
and in order to provide a comprehensive comparison, based on the CoMA
concept, two new optimal precoders are designed, one to minimize the
total transmit power and one to minimize the symbol error rate (SER)
for a given NOMA pair. In addition, the receiver complexity of CoMA
scheme is investigated and compared with conventional NOMA scheme.

For clarity we highlight the main contributions of this work as follows.
\begin{enumerate}
\item CoMA scheme is proposed and introduced for the first time to remove
receive SIC and reduce the complexity of user pairing NOMA MISO systems. 
\item New CoMA precoder that minimizes the transmit power for a given system
performance is designed. 
\item We further adapt CoMA concept to design new precoder that is able
to minimize the system error rate subject to total power constraint.
\item The complexity analysis of the proposed CoMA scheme is considered
and investigated.
\item The performance of CoMA scheme is compared with OMA and conventional
NOMA precoders.
\end{enumerate}
The results in this work show that CoMA scheme consumes much less
power than conventional NOMA and OMA techniques to achieve similar
target rates. In addition, CoMA scheme has lower error rate than OMA
and conventional NOMA schemes. Furthermore, our results confirm that
the new proposed CoMA scheme has very low computational receiver complexity
compared to conventional NOMA technique.

Next, Section \ref{sec:System-Model} describes the MU-MISO system
model. Section \ref{sec:Constructive-NOMA-(C-NOMA)}, introduces the
new proposed CoMA scheme. Section \ref{sec:Power-Minimization} considers
power minimization problems of CoMA and NOMA subject to QoS constraint.
Section \ref{sec:Error-Rate-minimization}, studies the error rate
minimization problems for CoMA and NOMA subject to total power constraint.
The computational receiver complexity of NOMA and CoMA are presented
in Section \ref{sec:Complexity}. Numerical results are presented
and discussed in Section \ref{sec:Numerical-Results}. Finally, Section
\ref{sec:Conclusions} concludes this paper.

\section{System Model\label{sec:System-Model}}

\begin{figure*}
\noindent \begin{centering}
\includegraphics[bb=280bp 90bp 760bp 500bp,clip,scale=0.6]{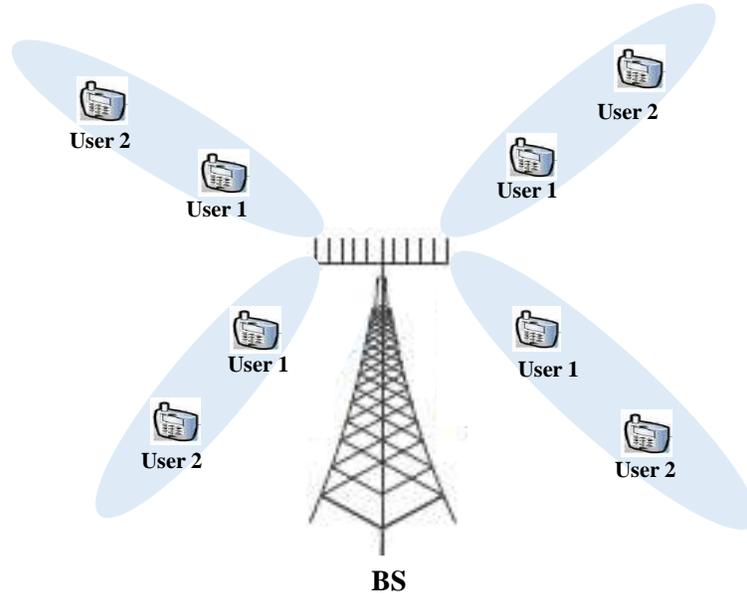}
\par\end{centering}

\protect\caption{\label{fig:sm}A multiuser NOMA system with $K$ pairs.}

\end{figure*}

We consider a down-link MU-MISO system, in which a BS equipped with
$N$ antennas transmits information signals to $2K$ single antenna
users using user-pairing NOMA technique \cite{twousers1,userpairs,pair1,pair2}.
In this system, each two users are paired to form a cluster, and hence,
there are $K$ pairs/clusters in the system as shown in Fig. \ref{fig:sm}.
Block fading channel model is assumed, in which each channel coefficient
includes both small scale fading and large scale fading. The $N\times1$
channel vector between the BS and user $i$, $i\in\left\{ 1,2\right\} $
in pair $k$, $k\in\left\{ 1,....,K\right\} $, is $\mathbf{h}_{k,i}\sim\mathcal{CN}\left(\text{0, }\mathbf{I}_{N}\sigma_{k,i}^{2}\right).$

Following the principle of NOMA, the BS broadcasts a superimposed
message of the two users in each pair. For pair $k$, the BS transmits
$\mathbf{x}_{k}=\mathbf{w}_{k,1}x_{k,1}+\mathbf{w}_{k,2}x_{k,2}$,
where $x_{k,1}$ and $x_{k,2}$ are the data symbols for user 1 $(u_{k,1})$
and user 2 $(u_{k,2})$ with unit variance, $\mathbf{w}_{k,i}$ is
the precoding vector of user $i$. In user-pairing NOMA scheme the
two users in each pair are ordered based on their CSI. Without loss
of generality, user 1, is assumed to has better channel than user
2, hence, the power allocated to user 2 should be higher than the
power allocated to user 1. The received signals at user 1 and user
2 in pair $k$ can be written as 

\begin{equation}
y_{u_{k,i}}=\mathbf{h}_{k,i}^{T}\stackrel[l=1]{2}{\sum}\mathbf{w}_{k,l}x_{k,l}+n_{u_{k,i}},\label{eq:1}
\end{equation}

\noindent where $n_{u_{k,i}}$ is the additive white Gaussian noise
(AWGN) at user $i$ with variance $\sigma_{u_{k,i}}^{2}$, $n_{u_{k,i}}\sim\mathcal{CN}\left(\text{0, }\sigma_{u_{k,i}}^{2}\right).$

\noindent Based on NOMA, the stronger user, user 1, adopts a SIC,
in which user 1 first detects user 2 signal, and then removes the
detected signal term from the received signal to decode its own message.
Thus, the received SINR at user 1 to detect user 2 signal, $x_{k,2}$,
can be written as,

\begin{equation}
\gamma_{x_{k,2}\rightarrow u_{k,1}}=\frac{\left|\mathbf{h}_{k,1}^{T}\mathbf{w}_{k,2}\right|^{2}}{\left|\mathbf{h}_{k,1}^{T}\mathbf{w}_{k,1}\right|^{2}+\sigma_{u_{k,1}}^{2}},\label{eq:3}
\end{equation}

\noindent The data rate for user 1 to detect user 2 signal, $R_{x_{k,2}\rightarrow u_{k,1}}$,
should be larger than the target rate of user 2 and thus $\gamma_{x_{k,2}\rightarrow u_{k,1}}$
should be higher than the target SINR at user 2 $(r_{2})$. The received
signal at user 1 after using SIC is given by 
\begin{equation}
y_{u_{k,1}}=\mathbf{h}_{k,1}^{T}\mathbf{w}_{k,1}x_{k,1}+\epsilon+n_{u_{k,1}},
\end{equation}

\noindent where $\epsilon$ is the SIC error with variance $\sigma_{\epsilon}^{2}$.
This error may occur due to incorrect detection of $x_{k,2}$ , incorrect
CSI knowledge, or incorrect knowledge of the power allocation at the
BS. 

\noindent Consequently, the received SINR at user 1 and user 2, to
detect $x_{k,1}$and $x_{k,2}$, respectively, can be written as

\begin{equation}
\gamma_{u_{k,1}}=\frac{\left|\mathbf{h}_{k,1}^{T}\mathbf{w}_{k,1}\right|^{2}}{\sigma_{\epsilon}^{2}+\sigma_{u_{k,1}}^{2}},\qquad\label{eq:5}
\end{equation}

\begin{equation}
\gamma_{u_{k,2}}=\frac{\left|\mathbf{h}_{k,2}^{T}\mathbf{w}_{k,2}\right|^{2}}{\left|\mathbf{h}_{k,2}^{T}\mathbf{w}_{k,1}\right|^{2}+\sigma_{u_{k,2}}^{2}}.\label{eq:6}
\end{equation}

However, as we have explained earlier, NOMA scheme suffers from a
key challenge. The need to perform SIC at the receiver on a symbol-by-symbol
level, i.e. for an LTE frame on the order of 0.1msec. This implicates: 
\begin{itemize}
\item Large complexity at the UE receiver that makes the practical application
challenging. 
\item Increased latency in the signal detection. 
\item Overheads in obtaining/feedback of CSI. 
\item Increased transmit power when the SIC errors increase.
\end{itemize}
In order to overcome all these challenging points CoMA technique is
proposed and presented in the next Section. 

$\,$

$\,$

$\,$

\section{\noindent Constructive NOMA (CoMA) Scheme\label{sec:Constructive-NOMA-(C-NOMA)}}

\begin{figure}[H]
\begin{centering}
\textcolor{blue}{\includegraphics[bb=250bp 150bp 790bp 500bp,clip,scale=0.6]{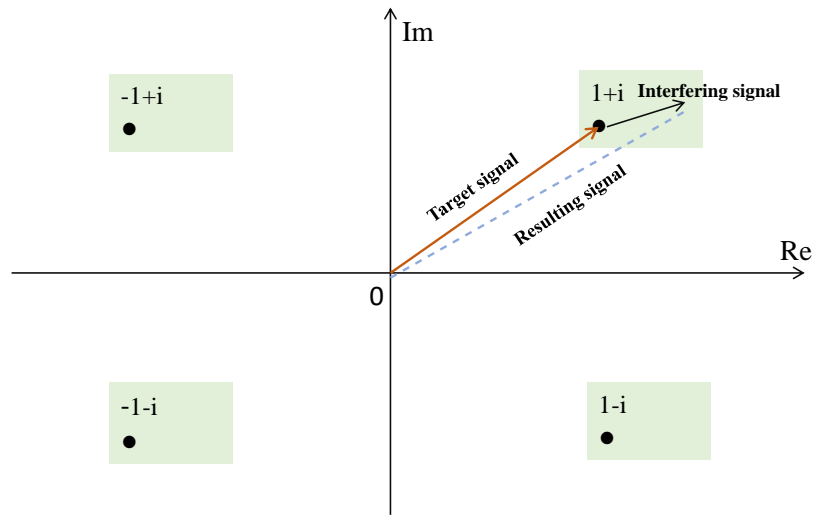}}
\par\end{centering}

\textcolor{blue}{\protect\caption{\label{fig:CI} The basic concept of CI in QPSK, the constructive
regions are represented by the green areas.}
}
\end{figure}

\begin{figure}[H]
\begin{centering}
\includegraphics[bb=0bp 120bp 960bp 540bp,clip,scale=0.55]{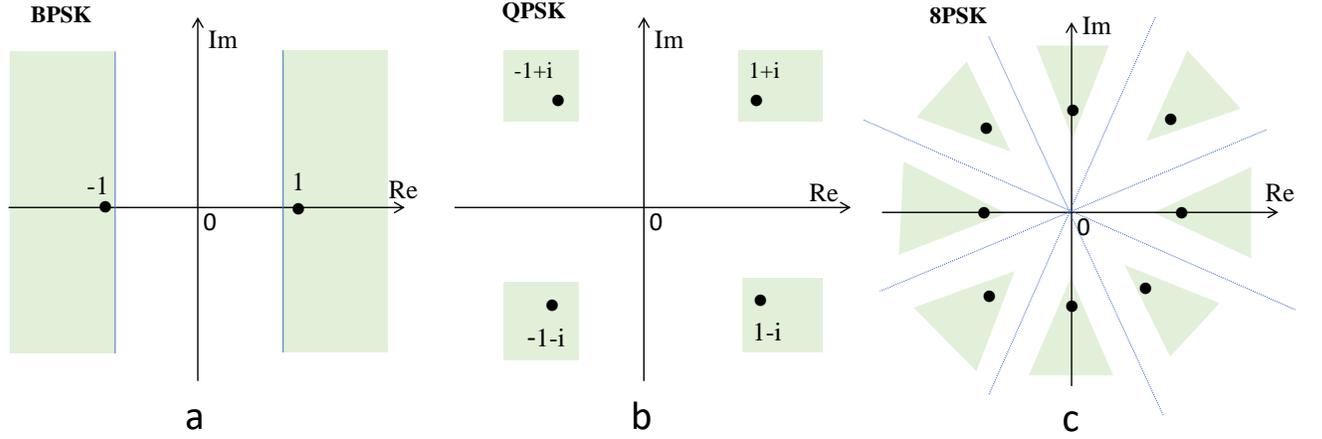}
\par\end{centering}

\textcolor{blue}{\protect\caption{\label{fig:CI-1} Constructive interference in a) BPSK, b) QPSK and
c) 8PSK, the constructive regions are represented by the green areas.}
}
\end{figure}

The main idea of CoMA scheme is to align the superimposed signals
that is known to the BS to increase the useful signal power at the
receiver. As we mentioned earlier, the interference signal is constructive
if it can move the received symbol towards the detection/constructive
region. The basic concept of CI precoding for QPSK constellation is
summarized in Fig. \ref{fig:CI}. Briefly, the constructive interference
pushes the received symbol deeper in the detection/constructive region,
this represents the green areas in the constellation of Fig. \ref{fig:CI},
and thus enhances the detection. Additionally, the constructive areas
for BPSK, QPSK and 8PSK are shown in Fig. \ref{fig:CI-1}. For more
details, the reader is referred to \cite{surveyAngLI} where the interference
exploitation scheme has been discussed and the practical implementation
of CI precoding has been presented. 

\textcolor{blue}{}

\textcolor{blue}{}

Therefore, following the CI principle \cite{surveyAngLI} the transmit
precoding can be designed to impose constructive interference to the
desired symbol. When the interference is aligned by means of precoding
vectors to overlap constructively with the signal of interest, all
interference contributes constructively to the useful signal and thus
the SINR expressions can be modified to take the constructive interference
into account. For the example of PSK signaling, the modulated symbols
of the users in pair $k$ can be expressed as $x_{k,i}$ = $x\, e^{j\phi_{i}}$,
where $x$ denotes the constant amplitude and $\phi_{i}$ is the phase.
Thus, the received signals at user 1 and user 2 presented in (\ref{eq:1})
can be represented as

\begin{equation}
y_{u_{k,i}}=\mathbf{h}_{k,i}^{T}\stackrel[l=1]{2}{\sum}\mathbf{w}_{k,l}e^{j\left(\phi_{l}-\phi_{i}\right)}x+n_{u_{k,i}},
\end{equation}

In CoMA scheme the interference at the strong user is designed to
be constructive to the desired symbol, thus the interference at user
1 contributes in the useful received signal power. Therefore, the
received SINRs at users 1 and 2 in pair $k$ are given, respectively,
by \cite{CI2,CI3}

\begin{equation}
\gamma_{u_{k,1}}=\frac{\left|\mathbf{h}_{k,1}^{T}\left(\mathbf{w}_{k,1}x_{k,1}+\mathbf{w}_{k,2}x_{k,2}\right)\right|^{2}}{\sigma_{u_{k,1}}^{2}},\qquad\label{eq:7}
\end{equation}

\begin{equation}
\gamma_{u_{k,2}}=\frac{\left|\mathbf{h}_{k,2}^{T}\mathbf{w}_{k,2}\right|^{2}}{\left|\mathbf{h}_{k,2}^{T}\mathbf{w}_{k,1}\right|^{2}+\sigma_{u_{k,2}}^{2}}.\label{eq:8}
\end{equation}

\begin{figure}[H]
\selectlanguage{american}%
\begin{centering}
\rule[0.1ex]{0.8\columnwidth}{0.8pt}
\par\end{centering}

\selectlanguage{english}%
\noindent \begin{centering}
\includegraphics[bb=100bp 50bp 940bp 450bp,clip,scale=0.55]{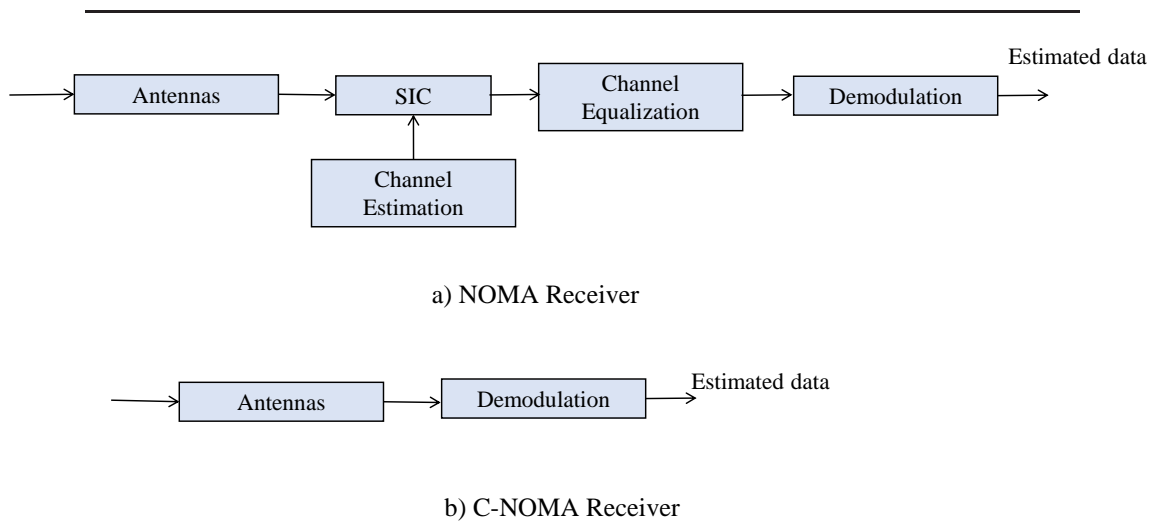}
\par\end{centering}

\selectlanguage{american}%
\begin{centering}
\rule[0.9ex]{0.8\columnwidth}{0.8pt}
\par\end{centering}

\selectlanguage{english}%
\protect\caption{\label{fig:2} Receivers of conventional and constructive NOMA schemes.}
\end{figure}

The block diagram of the user 1 receiver for conventional NOMA and
CoMA can be shown as in Fig. \ref{fig:2}. By comparing the two receivers
we can notice that by implementing CI, there is no need to use SIC
and channel equalization, and thus this simplifies the signal processing
procedure at the receiver side.

In the following Sections different precders are designed to minimize
the transmit power and the error rates of MU-MISO NOMA systems.

\section{\noindent Power Minimization \label{sec:Power-Minimization}}

In this Section we design prcoder vectors that minimize the total
transmission power subject to a quality-of-service (QoS) constrains.
For sake of comparison, CoMA and NOMA are considered in this Section.

\subsection{CoMA Precoding}

In this case we consider the power minimization problem for a given
pair with target SINR levels $r_{1},\, r_{2}$%
\footnote{From now and onward, for simplicity we omit the pair index $k$.%
}. As per the above classification and discussion, the optimization
problem can be formulated to take the constructive interference into
account in the power minimization problem. The total power consumption
in this case is $P=\left\Vert \stackrel[i=1]{2}{\sum}\mathbf{w}_{i}e^{j\left(\phi_{i}-\phi_{1}\right)}\right\Vert ^{2}.$
Accordingly and based on basic geometry of the constructive interference
regions \cite{CI2,CI3}, the optimization problem for $M$-PSK signaling
can be formulated as \cite{CI2,CI3} 

\begin{gather}
{\normalcolor \underset{\mathbf{w}_{i}\succeq\mathbf{0}}{\min}\:\left\Vert \stackrel[i=1]{2}{\sum}\mathbf{w}_{i}e^{j\left(\phi_{i}-\phi_{1}\right)}\right\Vert ^{2}}\nonumber \\
{\normalcolor s.t.\:\textbf{C1}:\left|\textrm{Im}\left(\mathbf{h}_{1}^{T}\stackrel[k=1]{2}{\sum}\mathbf{w}_{k}e^{j\left(\phi_{k}-\phi_{1}\right)}\right)\right|\leq\left(\textrm{Re}\left(\mathbf{h}_{1}^{T}\stackrel[k=1]{2}{\sum}\mathbf{w}_{k}e^{j\left(\phi_{k}-\phi_{1}\right)}\right)-\sqrt{r_{1}\sigma_{u1}^{2}}\right)\tan\theta}\nonumber \\
{\normalcolor \textbf{C2}:\left|\mathbf{h}_{2}^{T}\mathbf{w}_{2}\right|^{2}\geq\left(\left|\mathbf{h}_{2}^{T}\mathbf{w}_{1}\right|^{2}+\sigma_{u_{2}}^{2}\right)r_{2}}\label{eq:10-2}
\end{gather}

\noindent where $\theta=\pm\frac{\pi}{M}$. The first constraint in
(\ref{eq:10-2}), \textbf{C1,} is constructive interference constraint
for user 1 which is convex, please refer to \cite{CI2} for more details.
In addition, the second constraint in (\ref{eq:10-2}), \textbf{C2},
can be simplified using first order Taylor\textquoteright s approximation.
After applying the first-order Taylor expansion on $\bar{\mathbf{w}}_{i}$,
we can write

\begin{equation}
\left|\mathbf{h}_{2}^{T}\mathbf{w}_{2}\right|^{2}=2\textrm{Re}\left(\bar{\mathbf{w}}_{2}^{H}\mathbf{h}_{2}\mathbf{h}_{2}^{T}\mathbf{w}_{2}\right)-\textrm{Re}\left(\mathbf{\bar{w}}_{2}^{H}\mathbf{h}_{2}\mathbf{h}_{2}^{T}\mathbf{\bar{w}}_{2}\right)\label{eq:10-3}
\end{equation}

and

\begin{equation}
\left|\mathbf{h}_{2}^{T}\mathbf{w}_{1}\right|^{2}=2\textrm{Re}\left(\bar{\mathbf{w}}_{1}^{H}\mathbf{h}_{2}\mathbf{h}_{2}^{T}\mathbf{w}_{1}\right)-\textrm{Re}\left(\mathbf{\bar{w}}_{1}^{H}\mathbf{h}_{2}\mathbf{h}_{2}^{T}\mathbf{\bar{w}}_{1}\right)\label{eq:11}
\end{equation}

\noindent Therefore, (\ref{eq:10-2}) can be reformulated as 

\begin{gather}
\underset{\mathbf{w}_{i}\succeq\mathbf{0}}{\min}\:\left\Vert \stackrel[i=1]{2}{\sum}\mathbf{w}_{i}e^{j\left(\phi_{i}-\phi_{1}\right)}\right\Vert ^{2}\nonumber \\
s.t.\:\textbf{C1}:\left|\textrm{Im}\left(\mathbf{h}_{1}^{T}\stackrel[k=1]{2}{\sum}\mathbf{w}_{k}e^{j\left(\phi_{k}-\phi_{1}\right)}\right)\right|\leq\left(\textrm{Re}\left(\mathbf{h}_{1}^{T}\stackrel[k=1]{2}{\sum}\mathbf{w}_{k}e^{j\left(\phi_{k}-\phi_{1}\right)}\right)-\sqrt{r_{1}\sigma_{u1}^{2}}\right)\tan\theta\nonumber \\
\textbf{C2}:\,2\textrm{Re}\left(\bar{\mathbf{w}}_{2}^{H}\mathbf{h}_{2}\mathbf{h}_{2}^{T}\mathbf{w}_{2}\right)-\textrm{Re}\left(\mathbf{\bar{w}}_{2}^{H}\mathbf{h}_{2}\mathbf{h}_{2}^{T}\mathbf{\bar{w}}_{2}\right)\geq\nonumber \\
r_{2}\left(2\textrm{Re}\left(\bar{\mathbf{w}}_{1}^{H}\mathbf{h}_{2}\mathbf{h}_{2}^{T}\mathbf{w}_{1}\right)-\textrm{Re}\left(\mathbf{\bar{w}}_{1}^{H}\mathbf{h}_{2}\mathbf{h}_{2}^{T}\mathbf{\bar{w}}_{1}\right)\right)+\sigma_{u_{2}}^{2}r_{2}\label{eq:14}
\end{gather}

\noindent Finally, the all steps to solve (\ref{eq:14}) and find
the optimal precoding vectors using first-order Taylor expansion method
is presented in Algorithm 1.\textbf{}

\noindent 
\begin{algorithm}[H]
1: Set the maximum number of iterations $Q$. 

2: Randomly generate $\mathbf{\bar{w}}_{i}$. 

3: Repeat

4: Using CVX to solve (\ref{eq:14}) as $\mathbf{w}_{i}^{*}.$ 

5: Update $\mathbf{\bar{w}}_{i}=\mathbf{w}_{i}^{*}$ 

6: $q=q+1.$

7: Until $q=Q.$

8: Output $\mathbf{w}_{i}^{*}$ , $i\in K$.

\protect\caption{Iterative Algorithm for (\ref{eq:14}).}
\end{algorithm}

\subsection{Review: Conventional NOMA Precoding }

The total power consumption in conventional NOMA is $P=\stackrel[i=1]{2}{\sum}\left\Vert \mathbf{w}_{i}\right\Vert ^{2}$.
Consequently, from (\ref{eq:3}), (\ref{eq:5}) and (\ref{eq:6})
the power minimization problem can be formulated as

\begin{eqnarray}
\underset{\mathbf{w}_{i}}{\min}\: & \stackrel[i=1]{2}{\sum} & \left\Vert \mathbf{w}_{i}\right\Vert ^{2}\nonumber \\
s.t.\:\textbf{C1}:\gamma_{u_{1}} & \geq & r_{1}\nonumber \\
\textbf{C2}:\gamma_{u_{2}} & \geq & r_{2}\nonumber \\
\textbf{C3}:\gamma_{x_{2}\rightarrow u_{1}} & \geq & r_{2}\label{eq:13-2}
\end{eqnarray}

The constraint \textbf{C3} to ensure the successful SIC for the strong
user. The last expression in (\ref{eq:13-2}) can be presented in
more detailed formula as 

\begin{eqnarray}
\underset{\mathbf{w}_{i}}{\min}\: & \stackrel[i=1]{2}{\sum} & \left\Vert \mathbf{w}_{i}\right\Vert ^{2}\nonumber \\
s.t.\:\textbf{C1}:\left|\mathbf{h}_{1}^{T}\mathbf{w}_{1}\right|^{2} & \geq & \left(\sigma_{\epsilon}^{2}+\sigma_{u_{1}}^{2}\right)r_{1}\nonumber \\
\textbf{C2}:\left|\mathbf{h}_{2}^{T}\mathbf{w}_{2}\right|^{2} & \geq & \left(\left|\mathbf{h}_{2}^{T}\mathbf{w}_{1}\right|^{2}+\sigma_{u_{2}}^{2}\right)r_{2}\nonumber \\
\textbf{C3}:\left|\mathbf{h}_{1}^{T}\mathbf{w}_{2}\right|^{2} & \geq & \left(\left|\mathbf{h}_{1}^{T}\mathbf{w}_{1}\right|^{2}+\sigma_{u_{1}}^{2}\right)r_{2}\label{eq:13}
\end{eqnarray}

Semidefinite relaxation (SDR) can be used to obtain the optimal precodrs
in (\ref{eq:13}). The effectiveness of the SDR to solve this transmit
beamforming problem has been widely considered in literature \cite{sdr1,sdr2}.
 The problem in (\ref{eq:13}) has been investigated and considered
in details in \cite{newmain}, where the optimal and closed form solutions
have been provided. 

$\,$

$\,$

\section{\noindent SER Minimization \label{sec:Error-Rate-minimization}}

In this Section we design prcoder vectors to minimize the SER subject
to total transmission power constraint. For sake of comparison, CoMA
and NOMA are considered in this Section.

\subsection{CoMA Precoding }

In this Section we consider SER for the proposed COMA scheme. According
to \cite{ANgSER1} and \cite{ANgSER2}, the symbol error rate, SER,
can be expressed in the following form: 

\begin{equation}
\textrm{SER}=\frac{1}{K}\stackrel[k=1]{K}{\sum}f\left(\textrm{SNR}_{k}\right)
\end{equation}

where $f\left(SNR_{k}\right)$ is a function of user k\textquoteright s
SNR, which is determined by the modulation scheme. For example, if
QPSK modulation is employed, $f\left(SNR_{k}\right)$ can be further
expressed as

\begin{equation}
f\left(SNR_{k}\right)=\frac{1}{\sqrt{2\pi}}\stackrel[\textrm{SNR}_{k}]{\infty}{\int}e^{-\frac{1}{2}x^{2}}\: dx\label{eq:13-1}
\end{equation}

To minimize the SER of the proposed CoMA scheme, we construct the
following optimization problem:

\begin{gather}
\mathcal{P}_{1}:{\normalcolor \underset{\mathbf{w}_{1},\mathbf{w}_{2}}{\min}\:\underset{k}{\max}\left\{ \textrm{SER}_{k}\right\} }\nonumber \\
{\normalcolor s.t.\:\textrm{\textbf{C1}:}\left\Vert \stackrel[k=1]{2}{\sum}\mathbf{w}_{k}e^{j\left(\phi_{k}-\phi_{1}\right)}\right\Vert ^{2}<P}\nonumber \\
\textrm{\textbf{C2}:}\left|\textrm{Im}\left(\mathbf{h}_{1}^{T}\stackrel[k=1]{2}{\sum}\mathbf{w}_{k}e^{j\left(\phi_{k}-\phi_{1}\right)}\right)\right|\leq\left(\textrm{Re}\left(\mathbf{h}_{1}^{T}\stackrel[k=1]{2}{\sum}\mathbf{w}_{k}e^{j\left(\phi_{k}-\phi_{1}\right)}\right)-\sqrt{\textrm{SNR}_{1}\sigma_{u1}^{2}}\right)\tan\theta_{t},
\end{gather}

where \textbf{$P$ }is the total transmit power,\textbf{ C1} represents
the total transmit power budget and \textbf{C2} represents the constructive
interference constraint for user 1, respectively. The objective function
$\textrm{SER}_{k}$ is given by

\begin{equation}
\textrm{SER}=\frac{1}{\sqrt{2\pi}}\stackrel[\textrm{SNR}_{k}]{\infty}{\int}e^{-\frac{1}{2}x^{2}}\: dx
\end{equation}

Based on the SER expression in (\ref{eq:13-1}), the SER expression
for the two users in the system can be obtained as

\begin{equation}
\textrm{SER}_{1}=\frac{1}{\sqrt{2\pi}}\stackrel[\textrm{SNR}_{1}]{\infty}{\int}e^{-\frac{1}{2}x^{2}}\: dx
\end{equation}

\begin{equation}
\textrm{SER}_{2}=\frac{1}{\sqrt{2\pi}}\stackrel[\textrm{SNR}_{2}]{\infty}{\int}e^{-\frac{1}{2}x^{2}}\: dx
\end{equation}

According to the monotonicity that the SER decreases with the increase
of the received SNR, we transform the original optimization problem
into the following form:

\begin{gather}
\mathcal{P}_{2}:{\normalcolor \underset{\mathbf{w}_{1},\mathbf{w}_{2}}{\max}\:\underset{k}{\min}\left\{ \textrm{SNR}_{k}\right\} }\nonumber \\
{\normalcolor s.t.\:\textrm{\textbf{C1}:}\left\Vert \stackrel[k=1]{2}{\sum}\mathbf{w}_{k}e^{j\left(\phi_{k}-\phi_{1}\right)}\right\Vert ^{2}<P}\nonumber \\
\textrm{\textbf{C2}:}\left|\textrm{Im}\left(\mathbf{h}_{1}^{T}\stackrel[k=1]{2}{\sum}\mathbf{w}_{k}e^{j\left(\phi_{k}-\phi_{1}\right)}\right)\right|\leq\nonumber \\
\left(\textrm{Re}\left(\mathbf{h}_{1}^{T}\stackrel[k=1]{2}{\sum}\mathbf{w}_{k}e^{j\left(\phi_{k}-\phi_{1}\right)}\right)-\left|\left(\mathbf{h}_{1}^{T}\left(\mathbf{w}_{1}e^{j\phi_{1}}+\mathbf{w}_{2}e^{j\phi_{2}}\right)\right)\right|\right)\tan\theta_{t}\nonumber \\
\textrm{\textbf{C3}:}\textrm{Im}\left(\mathbf{h}_{2}^{T}\mathbf{w}_{2}\right)=0,\,\textrm{Re}\left(\mathbf{h}_{2}^{T}\mathbf{w}_{2}\right)\geqslant0
\end{gather}

where the additional constraint \textbf{C3} can guarantee that the
received symbol for user 2 lies in the correct decision region, while
the correct demodulation is guaranteed by the CI constraint, which
is well known. $\mathcal{P}_{2}$ transforms the original \textquotedblleft Max-
SER\textquotedblright{} minimization problem into a \textquotedblleft Min-SNR\textquotedblright{}
maximization problem. By further introducing an auxiliary variable
$\left(t\right)$, $\mathcal{P}_{2}$ can be further transformed into

\begin{gather}
\mathcal{P}_{3}:{\normalcolor \underset{\mathbf{w}_{1},\mathbf{w}_{2},t}{\max}\: t}\nonumber \\
{\normalcolor s.t.\:\textrm{\textbf{C1}:}\left\Vert \stackrel[k=1]{2}{\sum}\mathbf{w}_{k}e^{j\left(\phi_{k}-\phi_{1}\right)}\right\Vert ^{2}<P}\nonumber \\
\textrm{\textbf{C2}:}\left|\textrm{Im}\left(\mathbf{h}_{1}^{T}\stackrel[k=1]{2}{\sum}\mathbf{w}_{k}e^{j\left(\phi_{k}-\phi_{1}\right)}\right)\right|\leq\nonumber \\
\left(\textrm{Re}\left(\mathbf{h}_{1}^{T}\stackrel[k=1]{2}{\sum}\mathbf{w}_{k}e^{j\left(\phi_{k}-\phi_{1}\right)}\right)-\left|\left(\mathbf{h}_{1}^{T}\left(\mathbf{w}_{1}e^{j\phi_{1}}+\mathbf{w}_{2}e^{j\phi_{2}}\right)\right)\right|\right)\tan\theta_{t}\nonumber \\
\textrm{\textbf{C3}:}\textrm{Im}\left(\mathbf{h}_{2}^{T}\mathbf{w}_{2}\right)=0,\,\textrm{Re}\left(\mathbf{h}_{2}^{T}\mathbf{w}_{2}\right)\geqslant0\nonumber \\
\textrm{\textbf{C4}:}\left|\left(\mathbf{h}_{1}^{T}\left(\mathbf{w}_{1}e^{j\phi_{1}}+\mathbf{w}_{2}e^{j\phi_{2}}\right)\right)\right|^{2}\geqslant\sigma_{u1}^{2}t\nonumber \\
\textrm{\textbf{C5}:}\frac{\left|\mathbf{h}_{2}^{T}\mathbf{w}_{2}\right|^{2}}{\left|\mathbf{h}_{2}^{T}\mathbf{w}_{1}\right|^{2}+\sigma_{u_{2}}^{2}}\geqslant t
\end{gather}

where the auxiliary variable $t$ represents the minimum value of
the received SNR for the two CoMA users. In order to deal with the
non-convex fractional constraint \textbf{C5}, we propose to transform
its numerator into a concave function \cite{ANgSER3} by employing
the Taylor series expansion. More specifically, \textbf{C5} is re-expressed
as the following form

\begin{equation}
\frac{\mathcal{A}\left(\mathbf{w}_{2}\right)}{\mathcal{B}\left(\mathbf{w}_{1}\right)}\geqslant t,
\end{equation}

where the expressions for $\mathcal{A}\left(\mathbf{w}_{2}\right)$
and $\mathcal{B}\left(\mathbf{w}_{1}\right)$ are given by

\begin{eqnarray}
\mathcal{A}\left(\mathbf{w}_{2}\right) & = & 2\textrm{Re}\left(\bar{\mathbf{w}}_{2}^{H}\mathbf{h}_{2}\mathbf{h}_{2}^{T}\mathbf{w}_{2}\right)-\textrm{Re}\left(\mathbf{\bar{w}}_{2}^{H}\mathbf{h}_{2}\mathbf{h}_{2}^{T}\mathbf{\bar{w}}_{2}\right)\nonumber \\
\mathcal{B}\left(\mathbf{w}_{1}\right) & = & \left|\mathbf{h}_{2}^{T}\mathbf{w}_{1}\right|^{2}+\sigma_{u_{2}}^{2}
\end{eqnarray}

where $\mathbf{\bar{w}}_{2}$ represents the initial feasible point
of the Taylor series expansion. According to the Corollary 3 in \cite{ANgSER3},
we introduce another auxiliary variable $y$ to transform the nonconvex
constraint \textbf{C5} into a convex one:

\begin{equation}
2y\sqrt{\mathcal{A}\left(\mathbf{w}_{2}\right)}-y^{2}\mathcal{B}\left(\mathbf{w}_{1}\right)\geqslant t,
\end{equation}

Thus the corresponding optimization problem is further shown below:

\begin{gather}
\mathcal{P}_{4}:{\normalcolor \underset{\mathbf{w}_{1},\mathbf{w}_{2},t,y}{\max}\: t}\nonumber \\
{\normalcolor s.t.\:\textrm{\textbf{C1}:}\left\Vert \stackrel[k=1]{2}{\sum}\mathbf{w}_{k}e^{j\left(\phi_{k}-\phi_{1}\right)}\right\Vert ^{2}<P}\nonumber \\
\textrm{\textbf{C2}:}\left|\textrm{Im}\left(\mathbf{h}_{1}^{T}\stackrel[k=1]{2}{\sum}\mathbf{w}_{k}e^{j\left(\phi_{k}-\phi_{1}\right)}\right)\right|\leq\nonumber \\
\left(\textrm{Re}\left(\mathbf{h}_{1}^{T}\stackrel[k=1]{2}{\sum}\mathbf{w}_{k}e^{j\left(\phi_{k}-\phi_{1}\right)}\right)-\left|\left(\mathbf{h}_{1}^{T}\left(\mathbf{w}_{1}e^{j\phi_{1}}+\mathbf{w}_{2}e^{j\phi_{2}}\right)\right)\right|\right)\tan\theta_{t}\nonumber \\
\textrm{\textbf{C3}:}\textrm{Im}\left(\mathbf{h}_{2}^{T}\mathbf{w}_{2}\right)=0,\,\textrm{Re}\left(\mathbf{h}_{2}^{T}\mathbf{w}_{2}\right)\geqslant0\nonumber \\
\textrm{\textbf{C4}:}2\textrm{Re}\left(\left(\mathbf{\bar{w}}_{1}e^{j\phi_{1}}+\bar{\mathbf{w}}_{2}e^{j\phi_{2}}\right)^{H}\mathbf{h}_{1}\mathbf{h}_{1}^{T}\left(\mathbf{w}_{1}e^{j\phi_{1}}+\mathbf{w}_{2}e^{j\phi_{2}}\right)\right)-\nonumber \\
\textrm{Re}\left(\left(\mathbf{\bar{w}}_{1}e^{j\phi_{1}}+\bar{\mathbf{w}}_{2}e^{j\phi_{2}}\right)^{H}\mathbf{h}_{1}\mathbf{h}_{1}^{T}\left(\mathbf{\bar{w}}_{1}e^{j\phi_{1}}+\bar{\mathbf{w}}_{2}e^{j\phi_{2}}\right)\right)\geqslant\sigma_{u1}^{2}t\nonumber \\
\textrm{\textbf{C5}:}2y\sqrt{\mathcal{A}\left(\mathbf{w}_{2}\right)}-y^{2}\mathcal{B}\left(\mathbf{w}_{1}\right)\geqslant t,
\end{gather}

To solve $\mathcal{P}_{4}$, we adopt the block coordinate ascent
algorithm. Firstly, for given $\mathbf{w}_{1}$, $\mathbf{w}_{2}$
and $t$, the optimal value of $y^{*}$ can be obtained in a closed
form as

\begin{equation}
y^{*}=\frac{\sqrt{\mathcal{A}\left(\mathbf{w}_{2}\right)}}{\mathcal{B}\left(\mathbf{w}_{1}\right)}\geqslant t,\label{eq:26}
\end{equation}

Then, for given, $y^{*}$, $\mathbf{w}_{1}$, $\mathbf{w}_{2}$ and
$t$ can be obtained via solving $\mathcal{P}_{4}$ by substituting
$y^{*}$ into the constraint \textbf{C5} and solve the following optimization
problem $\mathcal{P}_{5}$:

\begin{gather}
\mathcal{P}_{5}:{\normalcolor \underset{\mathbf{w}_{1},\mathbf{w}_{2},t}{\max}\: t}\nonumber \\
{\normalcolor s.t.\:\textrm{\textbf{C1}:}\left\Vert \stackrel[k=1]{2}{\sum}\mathbf{w}_{k}e^{j\left(\phi_{k}-\phi_{1}\right)}\right\Vert ^{2}<P}\nonumber \\
\textrm{\textbf{C2}:}\left|\textrm{Im}\left(\mathbf{h}_{1}^{T}\stackrel[k=1]{2}{\sum}\mathbf{w}_{k}e^{j\left(\phi_{k}-\phi_{1}\right)}\right)\right|\leq\nonumber \\
\left(\textrm{Re}\left(\mathbf{h}_{1}^{T}\stackrel[k=1]{2}{\sum}\mathbf{w}_{k}e^{j\left(\phi_{k}-\phi_{1}\right)}\right)-\left|\left(\mathbf{h}_{1}^{T}\left(\mathbf{w}_{1}e^{j\phi_{1}}+\mathbf{w}_{2}e^{j\phi_{2}}\right)\right)\right|\right)\tan\theta_{t}\nonumber \\
\textrm{\textbf{C3}:}\textrm{Im}\left(\mathbf{h}_{2}^{T}\mathbf{w}_{2}\right)=0,\,\textrm{Re}\left(\mathbf{h}_{2}^{T}\mathbf{w}_{2}\right)\geqslant0\nonumber \\
\textrm{\textbf{C4}:}2\textrm{Re}\left(\left(\mathbf{\bar{w}}_{1}e^{j\phi_{1}}+\bar{\mathbf{w}}_{2}e^{j\phi_{2}}\right)^{H}\mathbf{h}_{1}\mathbf{h}_{1}^{T}\left(\mathbf{w}_{1}e^{j\phi_{1}}+\mathbf{w}_{2}e^{j\phi_{2}}\right)\right)-\nonumber \\
\textrm{Re}\left(\left(\mathbf{\bar{w}}_{1}e^{j\phi_{1}}+\bar{\mathbf{w}}_{2}e^{j\phi_{2}}\right)^{H}\mathbf{h}_{1}\mathbf{h}_{1}^{T}\left(\mathbf{\bar{w}}_{1}e^{j\phi_{1}}+\bar{\mathbf{w}}_{2}e^{j\phi_{2}}\right)\right)\geqslant\sigma_{u1}^{2}t\nonumber \\
\textrm{\textbf{C5}:}2y^{*}\sqrt{\mathcal{A}\left(\mathbf{w}_{2}\right)}-y^{*2}\mathcal{B}\left(\mathbf{w}_{1}\right)\geqslant t,
\end{gather}

It has been shown that the iterative algorithm converges within only
a few iterations. For clarity, we summarize the algorithm in Algorithm
2 below.

\noindent 
\begin{algorithm}[H]
Initialization : $\mathbf{\bar{w}}_{1}=0$, $\mathbf{\bar{w}}_{2}=0$,
$t=0$ 

Repeat 

Update $y$ based on (\ref{eq:26}); 

Update $\mathbf{w}_{1}$, $\mathbf{w}_{2}$ and $t$ by solving $\mathcal{P}_{5}$

Until Convergence

\protect\caption{Block Coordinate Ascent Algorithm for solving $\mathcal{P}_{5}$.}
\end{algorithm}

\textbf{}

\subsection{Review: Conventional NOMA Precoding }

Similarly, in order to minimize the SER of conventional NOMA scheme,
we can consider the following optimization problem:

\begin{gather}
\mathcal{P}_{1}:{\normalcolor \underset{\mathbf{w}_{1},\mathbf{w}_{2}}{\max}\:\underset{k}{\min}\left\{ \textrm{SNR}_{k}\right\} }\nonumber \\
{\normalcolor s.t.\:\textrm{\textbf{C1}:}\left\Vert \mathbf{w}_{1}\right\Vert ^{2}+\left\Vert \mathbf{w}_{2}\right\Vert ^{2}\leq P}\nonumber \\
\textbf{C2}:\gamma_{x_{2}\rightarrow u_{1}}\geq r_{2}
\end{gather}

The constraint \textbf{C2} to ensure the successful SIC for the strong
user. To deal with the non-convex objective function, an auxiliary
variable $t$ is introduced to equivalently convert the original problem
$\mathcal{P}_{1}$ into a new problem as follows

\begin{gather}
\mathcal{P}_{2}:{\normalcolor \underset{\mathbf{w}_{1},\mathbf{w}_{2},t}{\max}\: t}\nonumber \\
{\normalcolor s.t.\:\textrm{\textbf{C1}:}\left\Vert \mathbf{w}_{1}\right\Vert ^{2}+\left\Vert \mathbf{w}_{2}\right\Vert ^{2}\leq P}\nonumber \\
\textbf{C2}:\left|\mathbf{h}_{1}^{T}\mathbf{w}_{2}\right|^{2}\geq\left(\left|\mathbf{h}_{1}^{T}\mathbf{w}_{1}\right|^{2}+\sigma_{u_{1}}^{2}\right)r_{2}\nonumber \\
\textrm{\textbf{C3}:}\left|\mathbf{h}_{2}^{T}\mathbf{w}_{2}\right|^{2}\geqslant t\left|\mathbf{h}_{2}^{T}\mathbf{w}_{1}\right|^{2}+t\sigma_{u_{2}}^{2}
\end{gather}

Note that the objective function and the constraint \textbf{C1} in
$\mathcal{P}_{2}$ are convex, and the challenge is only in the constraints
\textbf{C2} and \textbf{C3}. However, we would like to mention that,
similar problem has been considered and solved in the literature using
bisection-based method, we refer the reader to \cite{serNOMA1,serNOMA2,SERNOMA3,SERNOMA4}
for more details.

\section{Receiver Complexity\label{sec:Complexity}}

In this section we focus our attention on the receiver complexity,
which impacts the user equipment where the computational resources
are scarce. To compare the computational receiver complexity of CoMA
with conventional NOMA, we provide here the complexity analysis for
the detection process. In the complexity analysis, the number of complex
operations is used as the complexity metric. The complexity of SIC
can be divided into two parts: decoding and subtraction. 

For classical NOMA, the weak user needs to detect its signal, while
the powerful user, first detects the weak user's signal, then subtracts
it from the received signal, before finally detecting its own signal.
Assuming an ML detector, the complexity of NOMA for pair $k$ can
be obtained as \cite{complexity1,complexity2}

\begin{equation}
\mathcal{C}_{NOMA-pair,k}=\stackrel[i=1]{2}{\sum}\underset{ML\textrm{ detection}}{\underbrace{\left(4NM_{k}+2M_{k}^{N}\right)}}\times\left(2-i+1\right)+\underset{\textrm{Subtraction}}{\underbrace{\mathcal{O}\left(M_{k}^{2}\right)}}
\end{equation}

\noindent where $M_{k}$ is the modulation order of pair $k$. The
total complexity of NOMA for all pairs can be written as

\begin{equation}
\mathcal{C}_{NOMA}=\stackrel[k=1]{K}{\sum}\left(\stackrel[i=1]{2}{\sum}\underset{ML\textrm{ detection}}{\underbrace{\left(4NM_{k}+2M_{k}^{N}\right)}}\times\left(2-i+1\right)+\underset{\textrm{Subtraction}}{\underbrace{\mathcal{O}\left(M_{k}^{2}\right)}}\right)
\end{equation}

\noindent where $K$ is number of pairs. 

\noindent On the other hand, for CoMA, the weak user needs to apply
classical detection for its signal, while the powerful user detects
its signal without the need to remove the interference. Accordingly,
the complexity of CoMA for pair $k$ can be written as \cite{complexity1,complexity2}

\begin{equation}
\mathcal{C}_{CI-NOMA-pair,k}=\underset{ML\textrm{ detection}}{\underbrace{\left(4NM_{k}+2M_{k}^{N}\right)}}+\underset{CI\textrm{ detection}}{\underbrace{D_{k}\left(M_{k}\right)}}
\end{equation}

\noindent where $D_{k}\left(M_{k}\right)$ is the complexity of decision
operation upon the received signal for pair $k$ which depends on
the modulation order.  The total complexity of CoMA for all pairs
can be obtained as

\begin{equation}
\mathcal{C}_{CI-NOMA}=\stackrel[k=1]{K}{\sum}\left(\underset{ML\textrm{ detection}}{\underbrace{\left(4NM_{k}+2M_{k}^{N}\right)}}+\underset{CI\textrm{ detection}}{\underbrace{D_{k}\left(M_{k}\right)}}\right).
\end{equation}

\section{Numerical Results\label{sec:Numerical-Results}}

To evaluate the performance of the proposed CoMA technique, in this
Section several numerical results for CoMA are presented and compared
with conventional NOMA and OMA using Monte Carlo simulations. In these
results we assume the users have same noise variance, $\sigma_{u_{1}}^{2}=\sigma_{u_{2}}^{2}=\sigma^{2}$.

\begin{figure*}
\noindent \begin{centering}
\subfloat[\label{fig:1a}Power consumption versus number of BS antennas when
$\left(r_{1},r_{2},\sigma_{1},\sigma_{2}\right)=(1,1,2,1).$ ]{\noindent \begin{centering}
\includegraphics[scale=0.7]{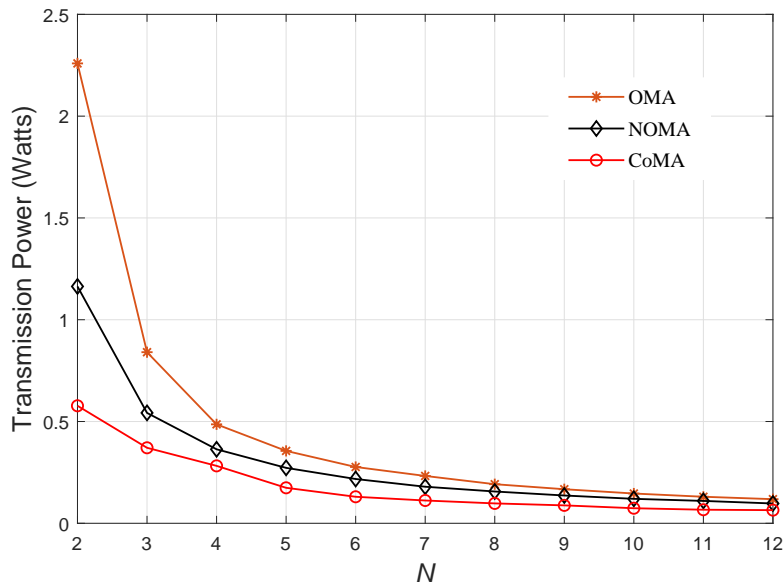}
\par\end{centering}

}
\par\end{centering}

\noindent \begin{centering}
\subfloat[\label{fig:1b} Power consumption versus number of BS antennas when
$\left(r_{1},r_{2},\sigma_{1},\sigma_{2}\right)=(1,1,3,1).$ ]{\noindent \begin{centering}
\includegraphics[scale=0.7]{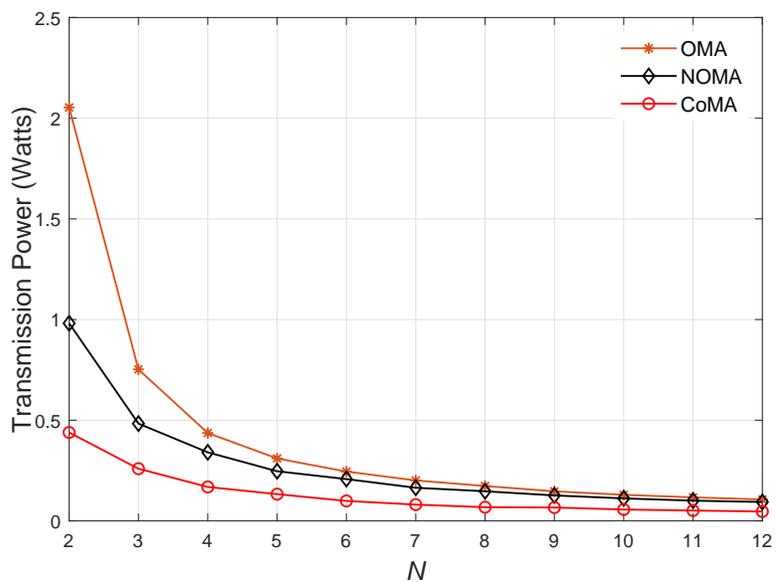}
\par\end{centering}

}
\par\end{centering}

\protect\caption{\label{fig:1-1}Power consumption for OMA, NOMA and CoMA versus number
of BS antennas for different channel variance.}
\end{figure*}

\begin{figure*}
\noindent \begin{centering}
\subfloat[\label{fig:1c} Power consumption versus number of BS antennas when
$\left(r_{1},r_{2},\sigma_{1},\sigma_{2}\right)=(1,3,2,1).$ ]{\noindent \begin{centering}
\includegraphics[scale=0.7]{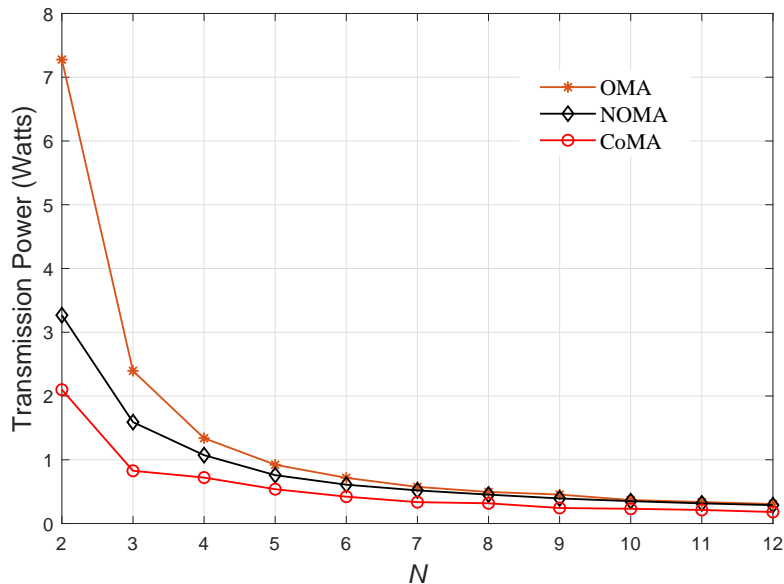}
\par\end{centering}

}
\par\end{centering}

\noindent \begin{centering}
\subfloat[\label{fig:1d} Power consumption versus number of BS antennas when
$\left(r_{1},r_{2},\sigma_{1},\sigma_{2}\right)=(3,1,2,1).$ ]{\noindent \begin{centering}
\includegraphics[scale=0.7]{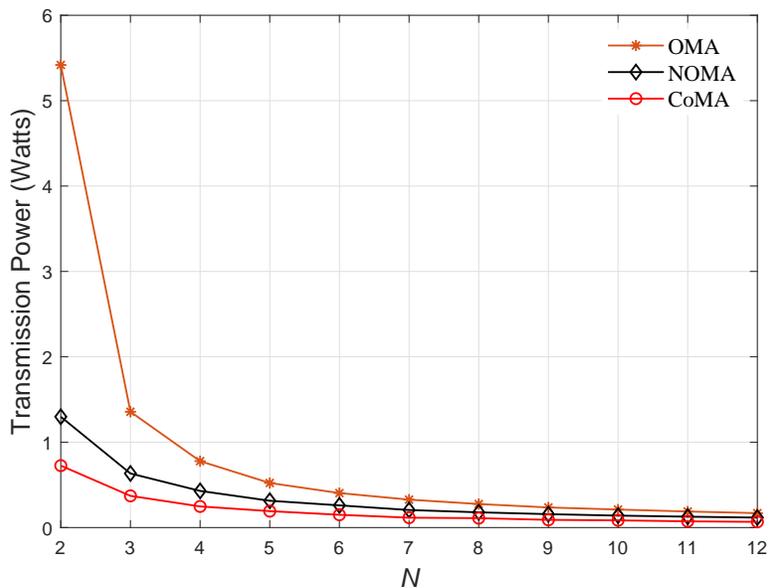}
\par\end{centering}

}
\par\end{centering}

\protect\caption{\label{fig:fig2new}Power consumption for OMA, NOMA and CoMA versus
number of BS antennas for different target rates.}

\end{figure*}

To measure the performance of CoMA technique in terms of total power
consumption for MU-MISO systems, in Fig. \ref{fig:1-1} we plot the
power consumption versus number of BS antennas $N$ for OMA, conventional
NOMA and CoMA with QPSK signaling using the power minimization approaches
in (\ref{eq:13}), and (\ref{eq:14}). The case when $\left(r_{1},r_{2},\sigma_{1},\sigma_{2}\right)=(1,1,2,1)$
is presented in Fig. \ref{fig:1a} and when $\left(r_{1},r_{2},\sigma_{1},\sigma_{2}\right)=(1,1,3,1)$
is shown in Fig.\ref{fig:1b}. Several observations can be extracted
from these results. Firstly it can be observed that, CoMA scheme has
a significant enhancement in terms of power consumption in comparison
with the conventional NOMA and OMA schemes. It is also noted that
the proposed CoMA scheme yields a significant performance gain in
the symmetric scenario when $N=2$. On the other hand, it is worth
pointing out that the difference between the considered schemes becomes
negligible when $N$ is large. Nevertheless, the complexity gains
of CoMA by removing the SIC operation persist. In addition, the total
transmission power decreases when the channel variance of user 1 increases
or if there is a notable disparity of channel strengths among users,
as shown in Figs. \ref{fig:1a} and \ref{fig:1b}. This is because
the strong user, user 1, in this case needs small power to achieve
its target rate, $r_{1}$. 

Furthermore, Fig. \ref{fig:fig2new} shows the power consumption for
OMA, NOMA and CoMA versus number of BS antennas for different target
rates. Fig. \ref{fig:1c} presents the case when $\left(r_{1},r_{2},\sigma_{1},\sigma_{2}\right)=(1,3,2,1)$
while Fig. \ref{fig:1d} shows the case when $\left(r_{1},r_{2},\sigma_{1},\sigma_{2}\right)=(3,1,2,1)$.
It can be clearly seen in these results that, increasing the target
rates of the two users leads to boost the transmission power, and
this increasing in the power is essential when the target rate of
the second user, $r_{2}$, is higher.

\begin{figure*}
\noindent \begin{centering}
\subfloat[\label{fig:7a}SER versus $P$ for QPSK when $N=2$. ]{\noindent \begin{centering}
\includegraphics[scale=0.65]{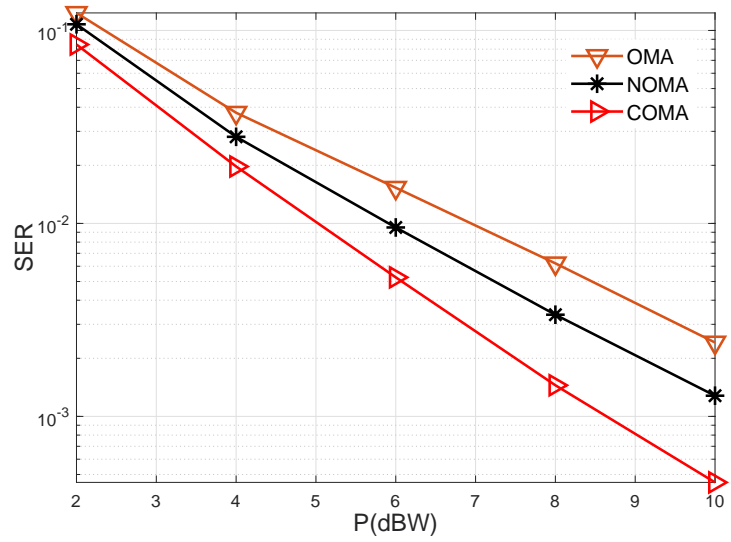}
\par\end{centering}

}
\par\end{centering}

\noindent \begin{centering}
\subfloat[\label{fig:7b}SER versus $P$ for QPSK when $N=4$.]{\noindent \begin{centering}
\includegraphics[scale=0.65]{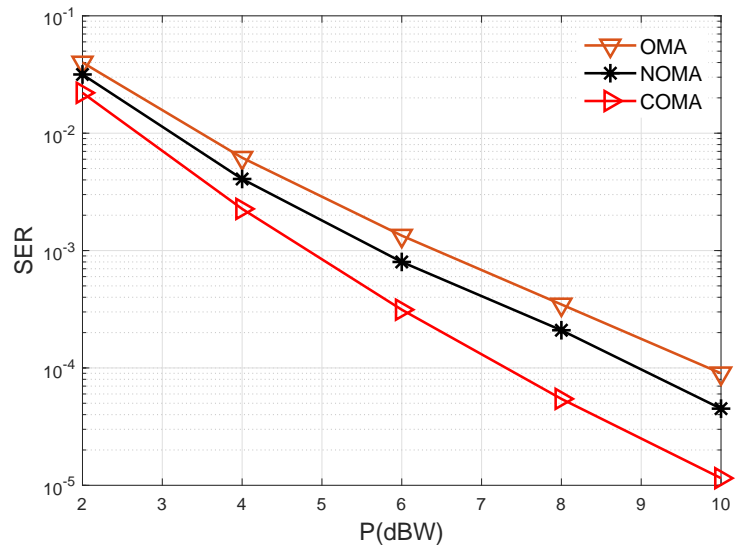}
\par\end{centering}

}
\par\end{centering}

\noindent \begin{centering}
\subfloat[\label{fig:7c}SER versus $P$ for 8PSK when $N=4$.]{\noindent \begin{centering}
\includegraphics[scale=0.65]{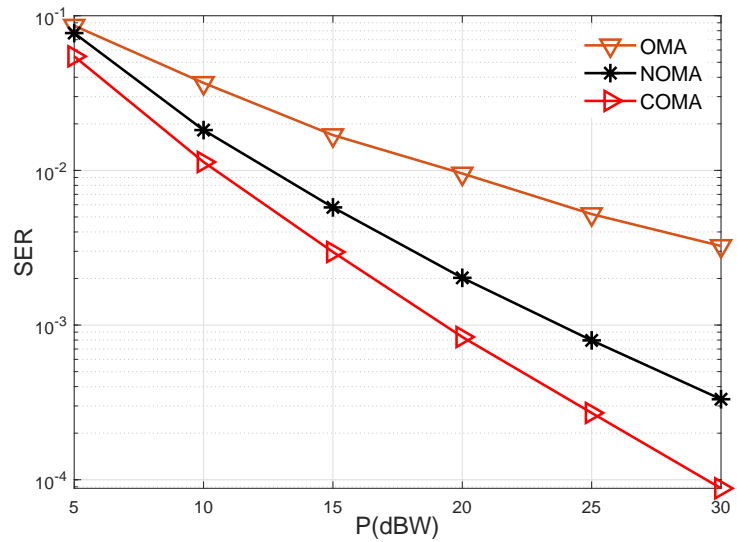}
\par\end{centering}

}
\par\end{centering}

\protect\caption{\label{fig:7}SER for OMA, NOMA and CoMA versus $P$ for different
number of antennas and modulation order.}
\end{figure*}

To evaluate the error rate performance of the proposed CoMA scheme,
Fig. \ref{fig:7} illustrates the SER versus the the total transmit
power, $P$, for OMA, conventional NOMA and COMA with different values
of number of BS antennas and modulation order. Fig. \ref{fig:7a}
and Fig. \ref{fig:7b} show the SER versus the total transmit power
when $N=2$ and $N=4$, respectively, for QPSK, while Fig. \ref{fig:7c}
represents the SER versus the transmit power when $N=4$ for 8PSK
scheme. Several interesting features can be noted in this figure.
Firstly, it is evident from these results that the SER reduces with
increasing the transmit power, and CoMA scheme always outperforms
OMA and conventional NOMA techniques in the all power levels. Looking
closer at Fig. \ref{fig:7a} and Fig. \ref{fig:7b} we can observe
that, increasing number of BS antennas reduces the SER, and the gain
attained by CoMA over conventional NOMA is almost fixed with the transmit
power. Finally, from Fig. \ref{fig:7b} and Fig. \ref{fig:7c} it
is clear that COMA has better performance than the other two schemes
and this superiority is major when the total transmit power is high.

\begin{figure}
\noindent \begin{centering}
\includegraphics[scale=0.7]{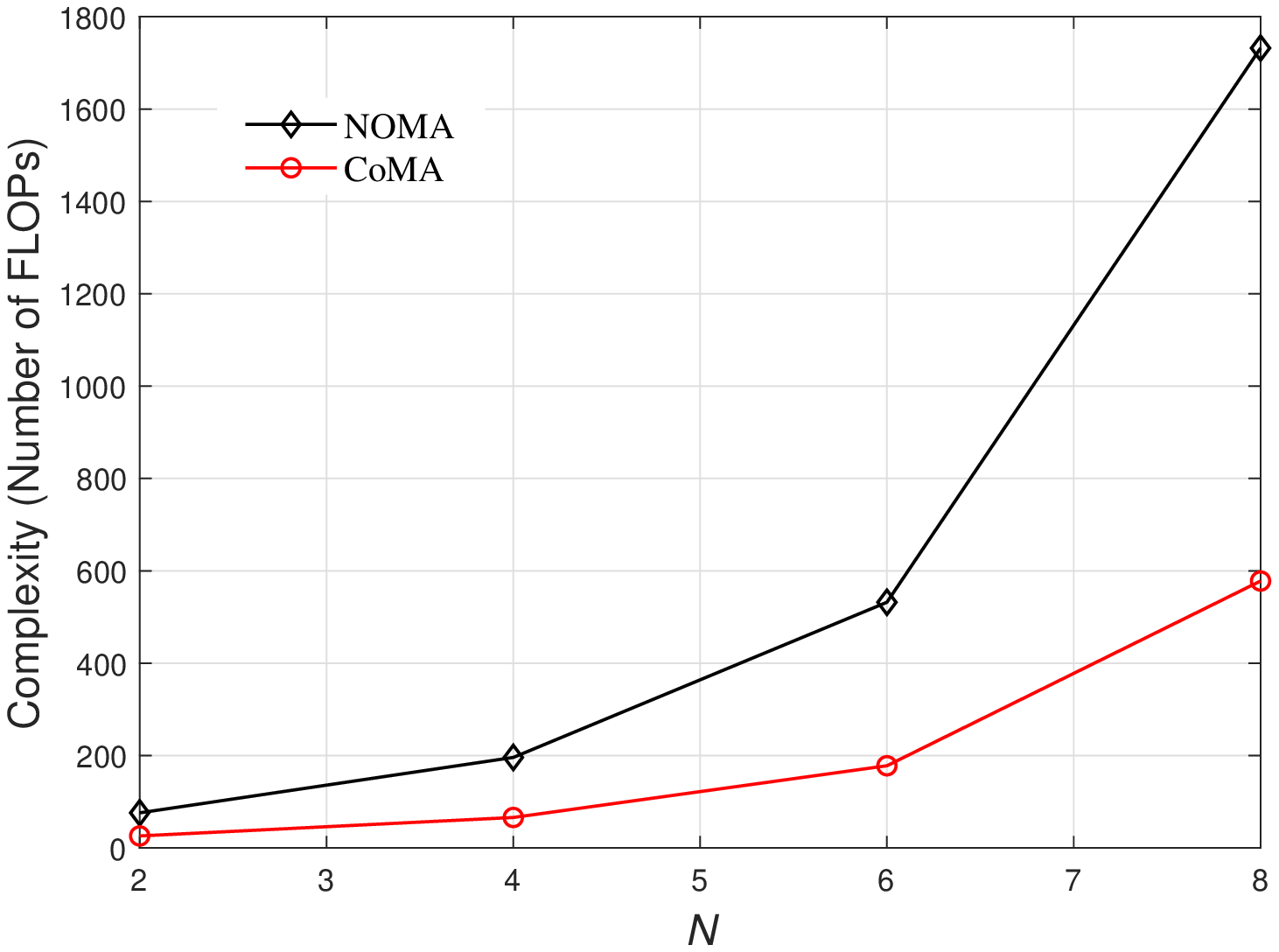}
\par\end{centering}

\protect\caption{\label{fig:4}Computational complexity versus number of BS antennas
for BPSK.}
\end{figure}

The computational receiver complexity of CoMA and conventional NOMA
versus number of BS antennas $N$ is presented in Fig. \ref{fig:4}.
It can be observed from these results that, CoMA substantially reduces
the computational complexity, which is desirable in hardware-limited
networks. In addition, the computational complexity gap between the
two schemes is much wider when number of BS antennas $N$ is high.

\begin{figure}
\noindent \begin{centering}
\includegraphics[scale=0.7]{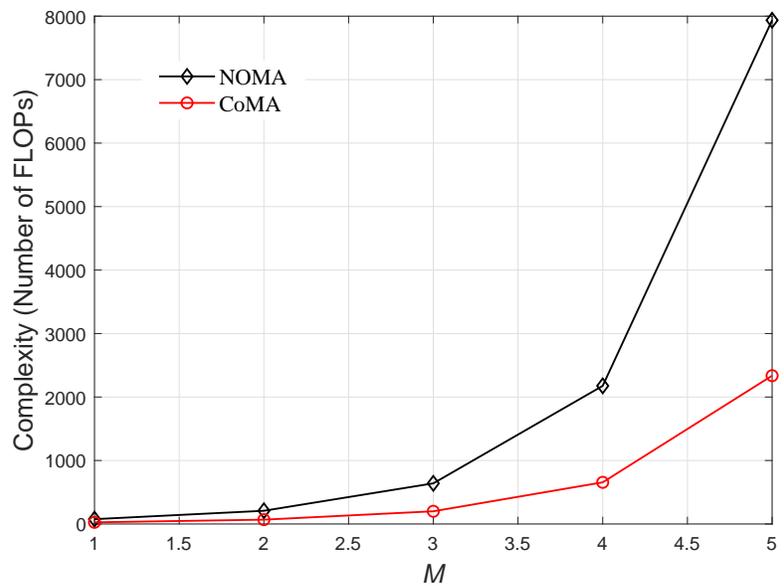}
\par\end{centering}

\protect\caption{\label{fig:5}Computational complexity versus modulation order.}
\end{figure}

Finally, Fig. \ref{fig:5} shows the computational receiver complexity
of CoMA and conventional NOMA versus the modulation order $M$ when
$N=2$. As shown in the figure, conventional NOMA scheme has higher
computational complexity than CoMA. In addition, the conventional
NOMA becomes computationally expensive for higher modulation orders.

\section{Conclusions\label{sec:Conclusions}}

In this paper a CoMA scheme for user pairing NOMA systems was proposed
and investigated. Firstly, for a given pair of users, the minimum
transmission power and the optimal precoding vectors of CoMA scheme
has been obtained. Then, optimal precoding vectors that minimizing
the symbol error rate subject to total power constrains for CoMA scheme
has been considered. Further, the complexity of CoMA has been studied
and compared with conventional NOMA scheme in terms of the total number
of complex operations. Simulation results have been provided to show
that, CoMA scheme consumes much less power than conventional NOMA
and OMA schemes to achieve similar target rates. In addition, CoMA
scheme produces lower error rate than conventional NOMA technique
over the all transmit power values. Furthermore, the proposed CoMA
scheme implicates very low computational receiver complexity compared
to conventional NOMA technique.

\section*{ACKNOWLEDGMENT }

This work is supported by the U.K. Engineering and Physical Sciences
Research Council (EPSRC) under grant EP/R007934/1.

\bibliographystyle{IEEEtran}
\bibliography{bbl}

\end{document}